\documentclass[a4paper]{jpconf}
\usepackage{graphicx}
\usepackage{epsfig}
\usepackage{epstopdf}
\usepackage{array}
\usepackage{siunitx}
\usepackage[usenames, dvipsnames]{color}
\usepackage{color}
\usepackage{subfig}
\usepackage{longtable}
\usepackage[utf8]{luainputenc}

\usepackage{xspace}
\usepackage{multicol}

\usepackage{makecell}
\usepackage{cite}

\newcommand{\Smilei}{{\sc Smilei}\xspace}
\usepackage{enumitem}

\begin{document}
\title{Single Domain Multiple Decompositions for Particle-in-Cell simulations}

\author{J. Derouillat$^1$, A. Beck$^2$}

\address{
$^1$ Maison de la Simulation, CEA, CNRS, Université Paris-Sud, UVSQ, Université Paris-Saclay, 91191 Gif-sur-Yvette, France\\
$^2$ Laboratoire Leprince-Ringuet – École polytechnique, CNRS-IN2P3, Palaiseau 91128, France
}
\ead{beck@llr.in2p3.fr}

\begin{abstract}
As a multi-purpose Particle-In-Cell (PIC) code, \Smilei gathers many different features in a single software.
Combining some of them is challenging.
In particular, spectral solvers and patch based load balancing have a priori non compatible requirements.
This paper introduces the Single Domain Multiple Decompositions (SDMD) method in order to address this issue.
To do so, different domain decompositions are used for fields and particles operations. 
This approach allows to keep small domains for particles, necessary for a good load balancing, while having large domains for the fields.
It proves beneficial in mitigating synchronization costs and gives the opportunity to introduce more paralellism in the PIC algorithm on top of providing
structures compatible with spectral solvers.

\end{abstract}

\section{Introduction}\label{intro}

Numerical simulations play an important role in the study of laser-plasma interaction and wakefield acceleration in particular.
For decades now, the core of these simulations has been the Particle-In-Cell (PIC) algorithm \cite{BirdsallLangdon1985}.
This algorithm's popularity can be explained by three main reasons.
1) It is physically relevant.
The PIC algorithm solves the Maxwell-Vlasov system, an accurate physical description for most plasma based accelerators.
2) It is conceptually very simple.
Electromagnetic fields evolve on a grid and the plasma phase-space is sampled by particle-like objects called macroparticles (MP).
3) It is relatively easy  and efficient to parallelize with a standard domain decomposition.
This last property is very important in the context of the massive growth of the number of processors in super-computers.

But the basic PIC simulation of laser wakefield acceleration (LWFA) experiments faces two major limitations.
It is affected by numerical artifacts \cite{Godfrey1974} and is very sensitive to resolution \cite{Lee2019}.
This makes PIC simulations rather costly in terms of computational resources and also limits the accuracy of the simulations.

Over the years, many improvements have been proposed in order to circumvent these limitations. 
On the one hand, new numerical methods intend to mitigate the numerical artifacts.
That category includes better techniques for solving Maxwell's equations and/or pushing MP as well as a variety of filters
applicable to fields or currents\cite{Godfrey2014}.
On the other hand, high performance computing (HPC) optimizations intend to improve the algorithm efficiency and scalability in order
to reach higher resolutions and therefore better accuracy.
This includes sorting techniques\cite{bowers2001,nakashima2015,smilei}, vectorization\cite{vincenti, smilei-vecto}, load balancing\cite{smilei,vay-loadbalance},
advanced programming models\cite{alpaka}, etc.

These improvements are done at the cost of an increased complexity which sometimes prevent their combination.
For instance, pseudo-spectral solvers\cite{Vay2013, Godfrey2015, Jalas2017} have interesting numerical properties but limits parallelism to large subdomains.
Conversely, for LWFA typical simulations, scalability and load balancing is significantly improved for small subdomains \cite{smilei}.
In this paper, this contradiction is addressed by using multiple decompositions of the simulation domain in order to fit the requirements of spectral solvers while keeping a good load balancing.
This ``Single Domain Multiple Decompositions'' (SDMD) technique uncouples fields related structures and MP related structures.
Doing so benefits to all fields related operations and can significantly improve performances.

Section \ref{method} describes the SDMD method, Section \ref{results} details different numerical configurations of a LWFA simulation and their performances.
Section \ref{discussion} discusses the obtained results and suggests further possible improvements.

\section{The SDMD method}\label{method}


\subsection{SDSD}\label{SDSD}
The standard domain decomposition technique in PIC codes will be referred to as ``Single Domain Single Decomposition'' (SDSD).
The domain is the physical volume simulated. 
The decomposition consists in splitting the domain into smaller subdomains which each represents only a fraction of the full domain.
This technique is suitable for distributed parallel computing because each subdomain can reside at different memory locations.

The SDSD method formally supports almost any number and shape of the subdomains.
Most dynamic load balancing techniques actually consist in changing their shape in time \cite{smilei, vay-loadbalance}.
The only mandatory property of the subdomains is that they form a partition of the domain: each point of the domain is in exactly one subdomain
and there are no empty subdomain.

In the case of the PIC algorithm, a subdomain represents a subvolume of the global simulation volume.
It stores all grid points and MP enclosed in that volume and communicates grid boundary values and exiting MP to its neighbor via operations called synchronizations.
A subdomain must be at least as large as a MP, i.e. contain at least as many points as the MP order of interpolation requires. 

\subsection{SDMD}\label{sdmd}

Following \ref{SDSD}, there are as many possible decompositions as there are partitions of the simulation domain.
SDSD picks a single one at any given time whereas the ``Single Domain Multiple Decompositions`` (SDMD) method consists in using several of them simultaneously.
``Single Domain'' here emphasizes that the same physical domain is addressed by different decompositions as opposed to other methods like
``Multi Levels Multi Domains'' \cite{MLMDparallel} in which the considered domains are physically different.

Having several decompositions available allows the use of the most suitable one for each part of the PIC algorithm.
A first approach is to use only two decompositions which formally differ only by the number of subdomains:
\begin{itemize}
\item The fine decomposition has many small subdomains called patches. It is dedicated to the MP operations.
\item The coarse decomposition has few large subdomains called regions. It is dedicated to the fields operations.
\end{itemize}

Patches and regions also differ by the information stored.
Patches are very much like an SDSD subdomain.
They hold all grid points and MP enclosed in their volume, whereas regions only need to store grid information.
Regions do not handle any MP at all.

This method has been implemented in the PIC code \Smilei.
Reference \cite{smilei} details the standard PIC loop and the patch based SDSD implementation of the code.
The addition of the coarse decomposition only slightly alters the algorithm with the two following additional steps:
\begin{enumerate}
 \item After current deposition on patches, current densities must be communicated from patches to regions.
 \item After solving Maxwell's equations on regions, electromagnetic fields must be communicated from regions to patches.
\end{enumerate}

Regions have two basic properties which limit the coarse decomposition possibilities:
\begin{enumerate}[label=\Alph*]
\item The regions should be as large as possible in order to optimize spectral solvers accuracy but also to minimize inter-region synchronizations.
In order to do so, the number of regions is set to the number of MPI processes used for the simulation.

\item Regions shapes should be constant in time and all the same for a good load balance.
It forces the use of rectangular shaped regions.
Note that in most cases, the number of MPI processes does not allow to have exactly the same shape for all regions and they might differ.
This difference should be kept as small as possible to ensure optimal load balance.
\end{enumerate}

In a SDMD simulation, each MPI process owns many patches and a single region.
Steps (i) and (ii) might involve non-local patches (which are not owned by the current MPI process).
The more overlap there is between the region and the local patches, the cheaper they are.
SDMD does not impose any patches or regions distributions between the MPI processes.
Nevertheless, these distributions should be done carefully and accommodate both a good load balance and a maximum overlap between patches and regions.

In \Smilei, patches are distributed along a Hilbert curve as a function of the local compute load of each patch\cite{smilei}.
Once the patches are distributed, regions are distributed according to which MPI rank owns the most patches overlapping the considered region.
This means that even if the regions are constant in time, their distribution between the MPI ranks is not and depends on the patches distribution.

\section{Results}\label{results}

The first and obvious benefit of the SDMD method is the ability to use spectral solvers in a patch based PIC code like \Smilei.
The solvers operate directly on the regions and are not aware of the patches.
But the SDMD method has a deeper impact on the simulation independently of the chosen Maxwell solver.
In this section we describe the numerical tests and comparisons that were carried out in order to analyze the effects of the SDMD method over the course of a standard
LWFA numerical simulation.

\subsection{Physical setup}

A laser of power $P=600$ TW, duration $\tau_L=25$ fs FWHM in intensity and wavelength $\lambda_0=800$ nm is focused at the entrance of a pre-ionized plasma.
The laser is modeled after the Apollon laser \cite{Cros2014}.
It is a perfectly top-hat transverse profile of diameter $D = 14$ cm focalized by a holed mirror.
The hole diameter is $d=31$ mm and the focal length is $f=3$ m.
The intensity profile in the focal plane is given by
\begin{equation}
 I(r) = \frac{4I_0}{\left(1-\epsilon^2\right)^2}\left(\frac{J_1(a)-\epsilon J_1(\epsilon a)}{a}\right)^2
\end{equation}

\noindent where $\epsilon=d/D$, $J_1$ is the Bessel function of the first kind, $a=r\pi D/\lambda_0 f$, $r$ is the distance to the optical axis and $I_0=P(1-\epsilon^2)\frac{\pi}{4}\left(\frac{D}{\lambda_0 f}\right)^2$.
The maximum intensity $I_{0} \approx 1.5\ 10^{20}\ \rm W/cm^2$ corresponds to a laser strength parameter $a_0=8.45$.

The plasma longitudinal density profile is a 500 microns long linear up ramp starting at the focal plane of the laser and reaching a plateau of $8.6\times10^{17}\ \rm cm^{-3}$.

\begin{figure}
\begin{center}
  
     \includegraphics[width=0.6\textwidth]{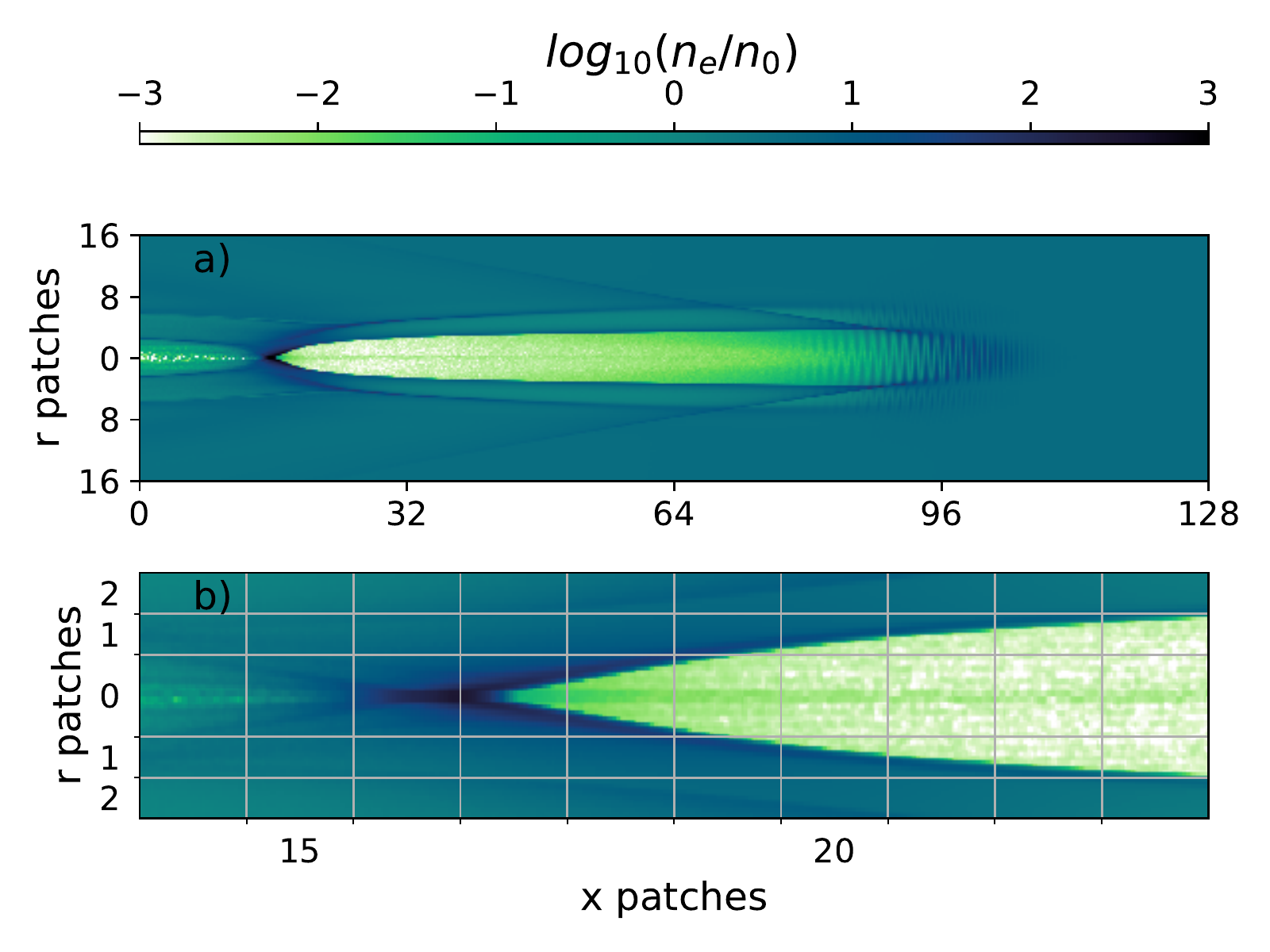}
\end{center}
 
 \caption{Logarithm of the electron density $n_e$ normalized to the plateau density $n_0$ in cylindrical geometry. 
 A typical blow out regime is established with a vast area empty of electrons in the wake of the laser and an extremely high density localized at the back of the so called ``bubble''.
 The axes show the $x$ and $r$ coordinates of the patches in a $128\times32$ patches case.
 Panel a) shows the central half of the simulation (patches 0 to 16 in $r$).
 Panel b) is a close up view of the hot spot at the origin of the imbalance in the second stage of the run.
 The lines delimit the patches boundaries.}
 \label{fig:blow_out}
\end{figure}

As the laser enters the plasma, the blow out regime, typical of many LWFA simulations, is quickly established as shown on figure \ref{fig:blow_out}.
The feature of interest for this study is the formation of the very high density spot at the back of the bubble highlighted on panel b) of figure \ref{fig:blow_out}.
This spot is responsible for most of the imbalance of the simulation which is studied in the following sections.

\subsection{Numerical setup}

The simulation runs in cylindrical geometry using an azimuthal decomposition\cite{Lifschitz2009} into 2 modes.
A standard finite difference scheme on a Yee lattice is used to solve Maxwell's equations.
$x$ is the longitudinal direction and $r$ the transverse one.
The domain is made of $4096\times448$ cells of dimension $1.6\times3\ (c/\omega_0)^2$ along $x$ and $r$ respectively where $\omega_0=2\pi c/\lambda_0$.
The time step is $0.159\ (1/\omega_0)$ and 11000 iterations are done.
Dynamic load balancing, if activated, is applied every 20 iterations.
A two-pass binomial filter\cite{BirdsallLangdon2004,Vay2015} is applied on the current densities at every iteration.
54 MP per cell are regularly positioned initially. 
Esirkepov current deposition scheme \cite{esirkepov:CPC2001} is used with a second order shape function (5 points stencil).
Figure \ref{fig:blow_out} displays the patches boundaries in a $128\times32$ patches case.
Note that in cylindrical geometry, cells and patches are toroidal and only those with radial coordinate $r=0$ are cylinders.
Also note that MP do not all have the same weight hence it is not straightforward to derive the compute load from the electronic density shown in figure \ref{fig:blow_out}.
Nevertheless, it is still a clear indication of a costly accumulation of MP in patches of radial coordinate $r=0$.
The rest of the paper illustrates and discusses how the simulation
performances are affected by it.
All simulations run on 50 KNL nodes hosted at `` Très Grand Centre de Calcul''.
They use a total of 200 MPI processes and 16 OpenMP threads per process.
\Smilei scalar operators are used\cite{smilei-vecto} so performances scale linearly with the number of MP.
Hyperthreading is not activated.

\subsection{Numerical measures}

The simulation is divided into two stages.
The first stage, referred to as the ``balanced stage'', goes from iterations  0 to 2000.
In that stage, the laser is still entering the domain, its wake is not formed yet and load imbalance is relatively weak.
Conversely, iterations from 2000 to 11000 are referred to as the ``unbalanced stage'' since load imbalance severely impacts this stage of the simulation.

Timers measure the time spent in different part of the PIC loop for each MPI ranks.
The loop is decomposed into:
\begin{itemize}
 \item Particles: interpolation, push, deposition.
 \item Fields: Maxwell solver and current density filter.
 \item Sync: synchronization of all fields, densities and MP.
 \item SyncDec: steps (i) and (ii) described in section \ref{sdmd}.
 \item MovWin: moving window operation.
\end{itemize}

\begin{figure}
\begin{center}
     \includegraphics[width=\textwidth]{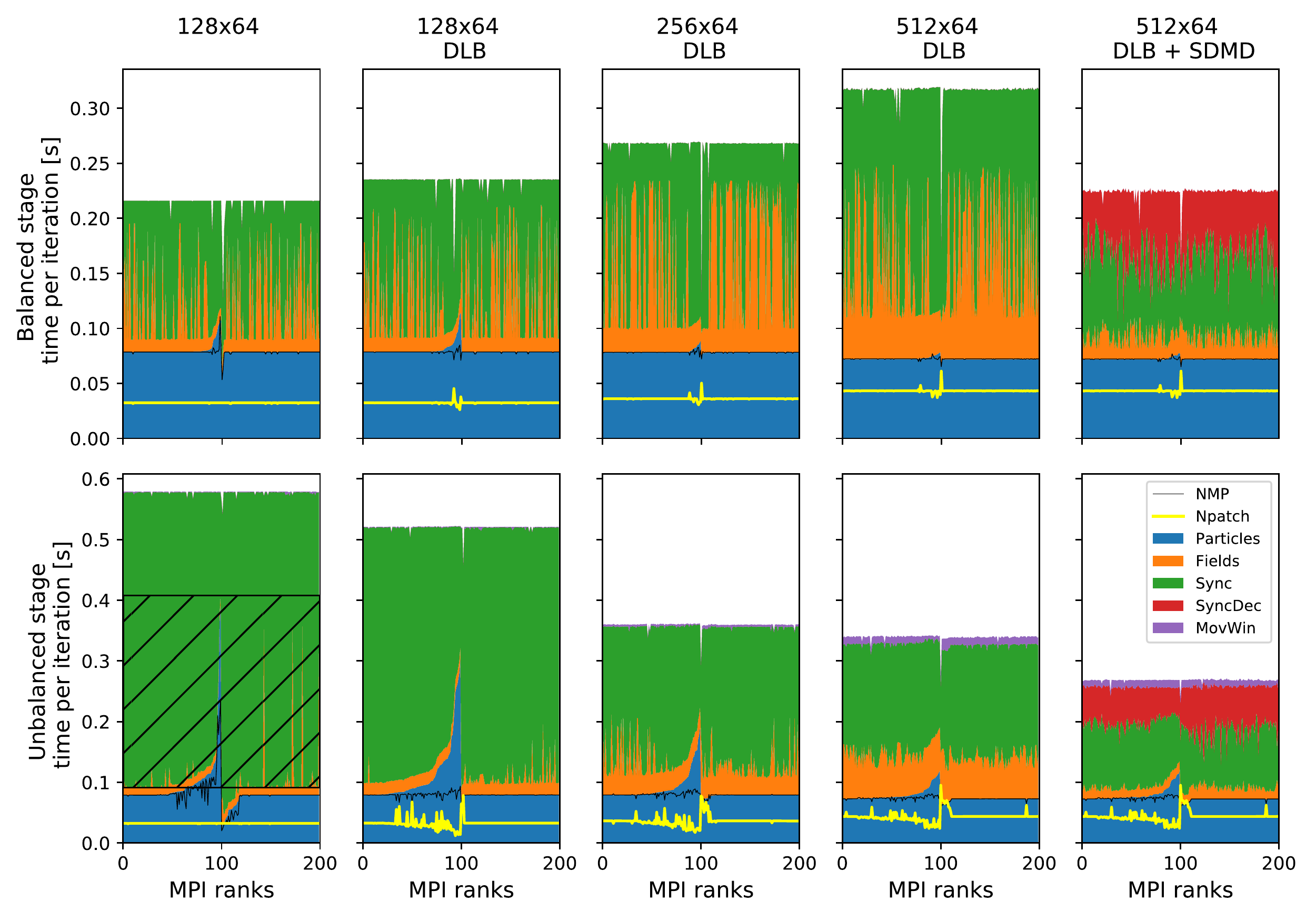}
\end{center}
 
 \caption{Average time per iteration spent in different parts of the PIC loop for each MPI rank.
  The first and second rows respectively show the timers of the balanced and unbalanced stages.
  Each column corresponds to a numerical configuration given at the top: number of patches in $x$ and $r$, dynamic load balancing (DLB) in columns 2 to 5 and SDMD in column 5.
 MPI ranks are sorted by increasing time spent in the particles timer and ``rolled'' such as maximum is at rank 99 and minimum at rank 100.
 The number of MP per MPI rank (NMP) is normalized so that the average NMP coincides with the average value of the particles timer.
 The number of patches per MPI rank (Npatch) is given in arbitrary units.
 The hatched area represents the idle time in the unbalanced stage without DLB.
 }
 \label{fig:perf_summary}
\end{figure}

Note that filters involve current densities synchronization between each pass which are accounted for in the ``Fields`` timer.
The number of MP and patches owned by each MPI rank is averaged in time over the corresponding stage.
Figure \ref{fig:perf_summary} gives an overview of these measures during both stages for different numerical configurations.

\subsection{Observations}

The first column of figure \ref{fig:perf_summary}, without DLB, shows an expected correlation between the time spent in treating particles and the number of MP.
The other columns show that particles are quite homogeneously distributed when DLB is activated but a significant imbalance in the particles timer is still measured (blue peak).
It is not correlated to the number of MP anymore and, as will be discussed in Sec. \ref{discussion}, this mainly follows from imbalance at the openMP thread level.

The imbalance in the particles timer appears to be quite critical.
In both balanced and unbalanced stages, and for any numerical configurations, the average time spent in particles is roughly the same ($\approx 0.08$ s per iteration).
Yet, the average total time spent per iteration in the unbalanced stage is roughly twice as much as in the balanced stage mostly because the time spent in Sync skyrockets.
This spectacular increase is a direct consequence of the imbalance in particles because, for the synchronization to start, each MPI rank has to wait for its neighbors to finish its particle operation.
And if A waits for B and B waits for C, then A waits for C.
Because of this transitivity of the dependencies, the expected total waiting time is eventually defined by the slowest MPI rank.
This naturally explains why numerical setups reducing particles imbalance reduces the Sync timer which in fact accounts for both
idle time and the actual synchronization operation itself.
This is illustrated by the hatched area on figure \ref{fig:perf_summary} which estimates the part of idle time in the Sync timer in the particular case of the unbalanced stage without DLB.
In this worst case scenario, idle time represents more than half of the total simulation time.

In the balanced stage, activating DLB and using smaller patches is actually detrimental to the performances.
Indeed, DLB implies additional operations and using more (and smaller) patches increases the cost of Fields and Sync operations.
Nevertheless, using SDMD almost completely cancels the fine grain decomposition overhead.

In contrast, in the unbalanced stage DLB significantly reduces the total simulation time.
As patches shrink, the DLB becomes more efficient but the cost of synchronizations increases.
Eventually, refining the decomposition from 256 to 512 patches along $x$, the gain in load balance is almost completely counterbalanced by the overhead.
Here again, this overhead can be avoided with the SDMD method which allows to fully benefit from the good load balancing provided by small patches.
Note that in that last case, the patch size is only $8\times7$ cells large which is hardly more than the particle shape function based on a 5 points stencil.

\section{Discussion}\label{discussion}
In both stages, the field operations and synchronizations are much more efficient on regions than on patches for two reasons.
First, these operations are completely balanced since each MPI process owns exactly one region whereas they each own a different number of patches.
Second, the total volume of communication is greatly reduced because the boundaries between the regions represent a much smaller fraction of the domain than the boundaries between the patches.

Looking at the number of patches on figure \ref{fig:perf_summary}, there is a correlation between longer time spent in particles (blue area) and lower number of patches
owned by the MPI processes (yellow line).
This means that the slowest MPI ranks are those owning fewer but more densely populated patches.

In \Smilei, patches are handled by different OpenMP threads.
The OpenMP dynamic scheduler automatically distributes the patches to idle threads.
Yet, if a MPI process owns too few patches and/or some of them are much more loaded than others, load imbalance might persist at the thread level and slow the whole process down.
This explains why load imbalance is still observed on processes with less patches even though MP look well distributed between MPI processes.
And the reason behind the better performances in the small patches configurations is a better distribution of the particles between OpenMP threads.



Scalability is limited by the OpenMP load balance which should still be improved.
As demonstrated on figure \ref{fig:perf_summary}, increasing the number of patches is one way to do so but the theoretical minimum limit to the patches size has almost
been reached already.
Further paralellization of the problem requires to break down a patch computation into several OpenMP tasks able to feed idle OpenMP threads.

\section{Conclusion}
Load imbalance is a major limitation of the LWFA simulations scalability.
The domain decomposition into as many subdomains as possible is necessary but difficult to achieve without loss of performances because of the synchronization overhead induced.
The SDMD method allows to efficiently use a domain decomposition with a close to the theoretical maximum number of patches while retaining reasonable synchronization times and the possibility to use pseudo-spectral solvers.
Further improvements of the parallelization will therefore have to go beyond domain decomposition and will probably involve task decomposition.

\ack
This work was granted access to the HPC resources of TGCC under the allocation 2019-A0060607484 madee by GENCI.
The authors also thank the engineers of the LLR HPC clusters for resources and help.

\section*{References}

\bibliography{Bibliography}
\bibliographystyle{iopart-num}

\end{document}